\documentclass[review,a4paper]{elsarticle}
\usepackage{amsmath}
\usepackage{color}
\usepackage{graphicx}
\usepackage{bm}
\usepackage{amssymb}
\usepackage{psfrag}
\usepackage[utf8]{inputenc}
\usepackage[toc,page]{appendix}
\usepackage{xcolor}
\usepackage{float}
\usepackage{url}
\usepackage{lineno}
\modulolinenumbers[5]

\usepackage{setspace}
\usepackage[section]{placeins} 

\newcommand{\cblue}{\textcolor{black}}

\journal{Communications in Nonlinear Science and Numerical Simulation}

\biboptions{numbers,sort&compress}

\begin{document}

\begin{frontmatter}

\title{Chaotic properties of an FIR filtered Hénon map}

\author{Vinícius S. Borges}
\fntext[mycorrespondingauthor]{Corresponding author}
\ead{viniciusb@usp.br}

\author{Magno T. M. Silva}
\fnref{myfootnote}\corref{mycorrespondingauthor}
\ead{magno.silva@usp.br}

\author{Marcio Eisencraft} \fnref{myfootnote}\corref{mycorrespondingauthor}
\ead{marcioft@usp.br}

\address{Escola Politécnica, \\University of São Paulo, Brazil}

\begin{abstract}
When chaotic signals are used in practical communication systems, it is essential to control and eventually limit the spectral bandwidth occupied by these signals. One way to achieve this goal is to insert a discrete-time filter into a nonlinear map that generates chaotic signals. However, this can completely change the dynamic properties of the original map. Considering this situation, this paper presents a series of numerical experiments aimed at obtaining the Lyapunov exponents of the signals generated by the two-dimensional Hénon map with a set of prototypical finite impulse response (FIR) filters added in the feedback loop. Our results show that the number of filter coefficients and the location of the zeros have a significant and complex impact on the behavior of the generated signals. Therefore, FIR filters should be carefully designed to preserve or suppress chaos in practical applications.
\end{abstract}


\end{frontmatter}


\section{Introduction}
\label{ch::introducao}

A chaotic signal has three main characteristics: it is bounded, aperiodic, and exhibits a sensitive dependence on initial conditions \cite{Alligood2000}. These properties often result in large bandwidth and have stimulated proposals for the application of chaotic signals in Telecommunications and Signal Processing, such as modulation \cite{kennedy2000digital, Souza2019,Baptista2021}, image encryption \cite{ismail2020novel,asgari2019novel,zhu2022stable}, 5G technology standards \cite{mohammed2022efficient,reena2021chaotic}, ultra-wideband communications \cite{eisencraft2010spectral}, among many others. 

Since transmission channels are usually limited in bandwidth \cite{Lathi2009}, knowing and controlling the spectrum of transmitted signals is a relevant problem when using chaotic signals in communication systems. With this in mind, the authors of  \cite{Eisencraft2009b} proposed  the use of a discrete-time finite impulse response (FIR) filter \cite{Oppenheim2009} in a chaotic signal generator as a way to control the spectral content of the generated signals. 

\cblue{Linear filters have been used  to denoise or detect chaotic signals \cite{Milosavljevic2020,Werner2017,Wang2011,Carroll1995,Butusov2018}. But in general, their output is not fed back into the generating system. Thus, they can be designed and analyzed in a way that is essentially separated from the chaos generator itself. However, in \cite{Eisencraft2009b, Fontes2016a} the filter is inserted into the chaos generator, and its output is fed back through the nonlinear system and through the filter itself. In this way, although the filter is actually linear, the resulting system is highly nonlinear, and the general behavior of the filtered chaos generator cannot be easily predicted.}

Subsequently, it was shown that insertion of this filter does not affect chaotic synchronization \cite{Fontes2016a}, which is essential for chaos-based communication. However, an important question has remained open: Since the inserted filters significantly alter the original map, are the generated signals still chaotic?

After the results of \cite{Fontes2016a}, the authors of \cite{borges2022f} analyzed the case of a two-coefficient FIR filter inserted in the feedback loop of a Hénon map. Using Lyapunov exponents, the authors have shown that, despite the simplicity of the filter considered, it can induce complicated phenomena such as cascades of bifurcations, coexistence of attractors, crises, and ``shrimps'' \cite{varga2020route,pati2020bifurcations}. Thus, the design of filters combined with the chaotic map can show dynamics that are difficult to predict and handle.

In this paper, we extend the research of \cite{borges2022f} by considering filters with a larger number of coefficients and investigating the effect of order and zero positions on the resulting dynamics. Due to the methodological challenges of considering many parameters in a dynamical system, we divide our analysis into five numerical experiments considering different prototypical FIR filters.  

The paper is organized as follows. The dynamical system under consideration is introduced in Section~\ref{section_henon_map}. The computation of its fixed points and a discussion of the stability of the orbits are presented in Section~\ref{section_fixed_point}. The numerical experiments with different FIR filters are presented in Section~\ref{section_experiments}.  Section~\ref{section_conclusion} closes the paper with the main conclusions.

\section{The filtered Hénon map}
\label{section_henon_map}

\begin{figure}[htb]
\centering
    \includegraphics[width=\textwidth]{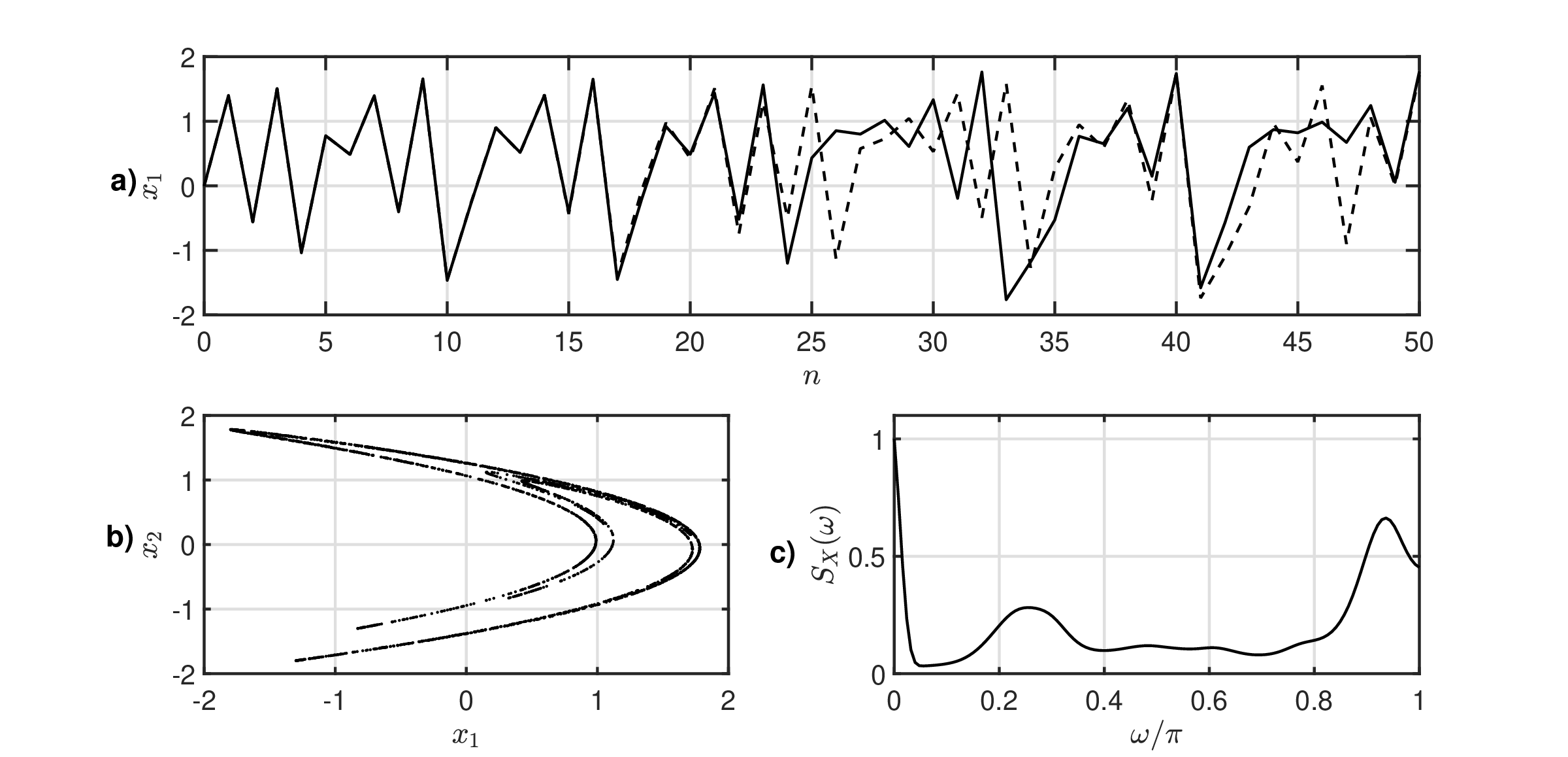}
    \caption{(a) Example of chaotic signals generated by (1) with $\alpha=1.4$ and $\beta = 0.3$. In solid line $x_1(0)=x_2(0)=0$ and in dashed line $x_1(0)=x_2(0)=0.0001$; \cblue{(b) phase space; and (c) power spectral density of the orbit with zero initial conditions}.}
    \label{fig::exemplohenon}
\end{figure}

The Hénon map is used as a paradigm to generate chaotic two-dimensional discrete-time signals \cite{Henon1969,henon1976two}. It can be expressed by
\begin{equation}
\begin{cases}
x_{1}(n+1) = \alpha - (x_{1}(n))^{2} + \beta x_{2}(n) \\
x_{2}(n+1) = x_{1}(n)
\end{cases},
\label{eq:henon}    
\end{equation}
where $\alpha$ and $\beta$ are real parameters and $x_1(n)$ and $x_2(n)$ are the state variables. In particular, in this paper, we consider $\alpha=1.4$ and $\beta = 0.3$ as in \cite{henon1976two}. These parameter values are known to produce chaotic signals, as shown in Figure~\ref{fig::exemplohenon}\cblue{(a)}, where the state variable $x_1(n)$ is plotted considering two close initial conditions. As can be seen, the two signals shown in Figure~\ref{fig::exemplohenon}\cblue{(a)} are bounded, aperiodic and present a sensitive dependence on initial conditions \cite{Alligood2000}. \cblue{Figures \ref{fig::exemplohenon}(b) and (c) show the phase space and the power spectral density of the orbit with initial condition equal to zero, respectively, highlighting the presence of a strange attractor and the aperiodic nature of the orbit.} 

To limit the spectral content of the signals generated by the Hénon map, a causal FIR filter \cite{Oppenheim2009} with $N_z$ zeros and coefficients $c_j$, $j=0, 1, \cdots, N_z$, $N_z\geq1$, $c_0\neq0$,  can be applied to $x_1(n)$, providing
\begin{equation}
x_{3}(n)=\sum_{j=0}^{N_{z}} c_{j} x_{1}(n-j).
\label{eq:FIR} 
\end{equation}
This filtered signal $x_3(n)$ is then fed back into the dynamic system, resulting in the \textit{filtered Hénon map} \cite{Eisencraft2009b,borges2022f}
\begin{equation}
\begin{cases}
x_{1}(n+1) = \alpha - (x_{3}(n))^{2} + \beta x_{2}(n) \\
x_{2}(n+1) = x_{1}(n)\\
x_{3}(n+1)=\sum_{j=0}^{N_{z}} c_{j} x_{1}(n-j+1)
\end{cases},
\label{eq:henonfilt} 
\end{equation}
where the state vector is $\mathbf{x}(n)\triangleq\left[x_1(n),x_2(n), x_3(n)\right]$.

In the analysis performed in \cite{borges2022f}, the simplest case $N_z=1$ was considered in detail, which already leads to complex phenomena, e.g., islands of periodicity in a sea of chaos, known as ``shrimps'' \cite{varga2020route,pati2020bifurcations}. Our goal here is to further extend the results presented there to more general FIR filters. The task is not straightforward due to the intrinsic nonlinearity of the system and the increase in dimensionality with $N_z$. 

To systematize the analysis, we will write the FIR filter in factored form and define its gain.
Thus, the transfer function $H(z)$ of the FIR filter \eqref{eq:FIR} can be written as \cite{Oppenheim2009}
\begin{equation}
H(z)\triangleq\sum_{j=0}^{N_z}c_jz^{-j}=c_0\prod_{k=1}^{N_z}\frac{(z-z_k)}{z},
\label{eq:fat1}
\end{equation}
where $z_k$, $k=1, 2,\cdots, N_z$ are the zeros of the filter, i.e., the roots of the polynomial $z^{N_z}H(z)$.
Defining the \emph{filter gain} as
\begin{equation}
G\triangleq H(1)=\sum_{j=0}^{N_z}c_j=c_0\prod_{k=1}^{N_z}{(1-z_k)}
\label{eq:gainG}
\end{equation}
and assuming that $z_k\neq 1$, $1\leq k\leq N_z$, 
it will be convenient for our analysis to express \eqref{eq:fat1} as
\begin{equation}
H(z)=G\prod_{k=1}^{N_z}\frac{(z-z_k)}{z(1-z_k)}.
\label{eq:fat2}
\end{equation}

We begin our analysis by studying the fixed points of \eqref{eq:henonfilt} and their stability in the following section. 

\section{Fixed points and their stability}
\label{section_fixed_point}

To compute the fixed points $\mathbf{p}=\left[p_1, p_2, p_3\right]$ of \eqref{eq:henonfilt}, we have to solve for 
\begin{equation}
\begin{cases}
p_{1}&=\alpha-(p_{3})^{2}+\beta p_{2} \\
p_{2}&=p_{1} \\
p_{3}&= p_{1} \sum_{j=0}^{N_{z}} c_{j}
\end{cases}.
\label{eq:fixedpoint}
\end{equation}
Using \eqref{eq:gainG} and replacing $p_2$ and $p_3$ in the first equation, we get
\begin{equation}
G^2p_1^2+(1-\beta) p_{1} - \alpha=0. 
\label{eq:fp}
\end{equation}
Thus, for $G\neq0$,
\begin{equation}
p_{1}^{+,-} = \frac{-(1-\beta)\pm \sqrt{(1-\beta)^{2} + 4\alpha G^{2}}}{2G^{2}}
\label{eq:pontofixop1}
\end{equation}
where $p_1^+>0$ and $p_1^-<0$. Replacing \eqref{eq:pontofixop1} in \eqref{eq:fixedpoint}, we obtain the fixed points
\begin{equation}
\mathbf{p}^+=\left[p_1^+, p_1^+, Gp_1^+\right]     
\end{equation}\label{eq:p+} and 
\begin{equation}
\mathbf{p}^-=\left[p_1^-, p_1^-, Gp_1^-\right].
\end{equation} 

\begin{figure}[htb]
\centering
    \includegraphics[width=\textwidth]{ 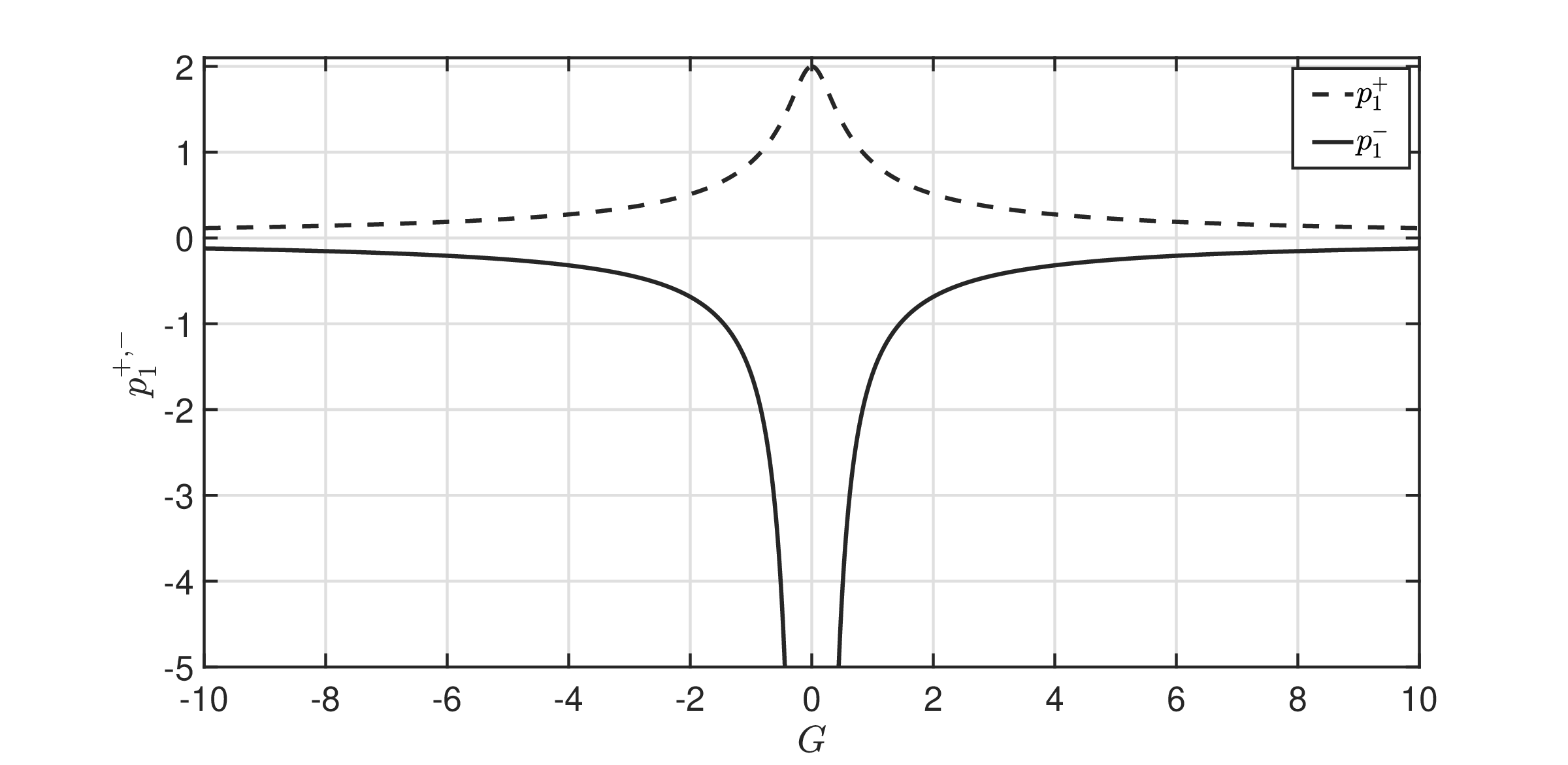}
    \caption{Plots of $p_1^+$ and $p_1^-$ of \eqref{eq:pontofixop1}. Note that $p_1^-$ diverges and $p_1^+\rightarrow p_0$ for $G=0$. }
    \label{fig:fixedpoint}
\end{figure}

Interestingly, fixed points depend only on the map parameters ($\alpha$ and $\beta$) and on the filter gain $G$, and not on the individual filter coefficients. 
The dependence of fixed points on the sum of the coefficients $G$ of \eqref{eq:gainG} agrees with the results of \cite{borges2022f} where a special case of \eqref{eq:pontofixop1} was obtained for $N_z=1$. 
We also note that for the particular case $G=0$,
\eqref{eq:fp} reduces to a linear equation, and there is only one fixed point at $\left[p_0,p_0,0\right]$ where 
\begin{equation}
p_0=\frac{\alpha}{1-\beta}=2,\label{eq:p0}\end{equation} 
for $\alpha=1.4$ and $\beta=0.3$, as we consider here. Figure \ref{fig:fixedpoint} shows $p_1^+$ and $p_1^-$ as a function of $G$. 

To determine the linear stability of the fixed points, we need to transform \eqref{eq:henonfilt} into a system of $N_z+1$ first-order difference equations. Then, it can be obtained from the absolute value of the eigenvalues of the associated Jacobian matrix $\mathbf{J}_{N_z}\left(\mathbf{x}(n)\right)$ evaluated at $\mathbf{p}^+$ and $\mathbf{p}^-$ \cite{Alligood2000,strogatz2018nonlinear}. The calculation of $\mathbf{J}_{N_z}\left(\mathbf{x}(n)\right)$ is presented in \ref{Apen:Jacobian}.

Although the coordinates of the fixed points depend only on $G$, our calculations in \ref{Apen:Jacobian} show that $\mathbf{J}_{N_z}\left(\mathbf{x}(n)\right)$, and consequently their linear stability, is more subtle, depending on the full set of filter coefficients. Our numerical analysis shows that $\mathbf{p}^-$ is always linearly unstable. This is not the case for $\mathbf{p}^+$. Therefore, in Section \ref{section_experiments}, we numerically investigate its stability by considering different prototype filters that lead to different sets of coefficients.
If $\mathbf{p}^+$ is unstable, the orbits of \eqref{eq:henonfilt} can converge to a periodic or chaotic attractor or diverge to infinity. To distinguish bounded orbits as periodic or chaotic, we numerically estimate their largest Lyapunov exponent $\Lambda$ using the tangent map technique \cite{Alligood2000}. An orbit with $\Lambda>0$ is considered a chaotic orbit, and an orbit with $\Lambda<0$ is considered to converge to a periodic or quasiperiodic attractor.

\section{Numerical Experiments}
\label{section_experiments}

In this section, a series of numerical experiments\footnote{\cblue{The codes used in all numerical experiments are avaiable at Github: \url{https://github.com/vinicius-s-borges/Chaotic-properties-of-an-FIR-filtered-Henon-map}}} are presented to illustrate the properties of the orbits of the filtered Hénon system \eqref{eq:henonfilt}. \cblue{All our experiments were performed with double precision, which we consider sufficient to avoid the influence of digital noise and truncation.} 

We consider some specific cases of the FIR filter \eqref{eq:fat2} with \textbf{(I)} $N_z$ zeros equally spaced in the unit circle; \textbf{(II)} a pair of complex conjugate zeros at different frequencies on the unit circle; \textbf{(III)} same as (II) but now with a third zero at $z=-1$; \textbf{(IV)} $N_z$ zeros at $z=-1$; and \textbf{(V)} zeros designed using Hamming windows so that the filter is lowpass \cite{Oppenheim2009}. These chosen cases can shed some light on the general dependence of the orbits of \eqref{eq:henonfilt} on $N_z$, $G$, and $z_k$ in \eqref{eq:fat2}.

In all experiments, the largest Lyapunov exponent $\Lambda$ was calculated considering $3\,000$ iterations of \eqref{eq:henonfilt}, eliminating the first 500. Furthermore, the largest Lyapunov exponent was averaged over 25 different initial conditions, uniformly chosen   in a sphere of radius $0.01$ centered on the fixed point $\textbf{p}^+$ of \eqref{eq:p+}. Our numerical experiments have shown that these quantities of initial conditions and iterations are sufficient to determine $\Lambda$ of the attractor with an accuracy higher than $10^{-2}$. 

We categorize each parameter set  according with the assimptotic properties of the orbits it can generate into four groups shown in different colors in the section figures: \textbf{(a)} convergence to $\mathbf{p}^+$ in black; \textbf{(b)} convergence to a periodic orbit in \textcolor{blue}{blue}; \textbf{(c)} chaotic orbits in \textcolor{red}{red}; and \textbf{(d)} 
 divergence to infinity in \textcolor{gray}{gray}.

A parameter set is of type (a) if the largest absolute value $\lambda$ of the eigenvalues of the Jacobian matrix $\mathbf{J}_{N_z}\left(\mathbf{p}^+\right)$ (see \ref{Apen:Jacobian})  is less than $1$, that is, $\lambda<1$. Type (b) is characterized for $\lambda>1$ and the generation of bounded orbits with the maximum Lyapunov exponent $\Lambda<0$. Type (c) also has $\lambda>1$ but $\Lambda>0$. Finally, if one of the random initial conditions tested generates an orbit that diverges in less than $3\,000$ iterations, it is considered type (d).

For cases (b) and (c), it is worth emphasizing that we are considering the average of the largest Lyapunov exponents calculated over a set of initial conditions. Thus, not all initial conditions necessarily produce the expected periodic or chaotic behavior. Our classification means that we have found orbits of this type for a given set of parameters. In some cases, it was even possible to detect the coexistence of periodic and chaotic attractors with large attraction basins, as in Experiment III, described in Section \ref{subsec:expIII}. This kind of complexity has also been reported in the simplified scenario of \cite{borges2022f}.

\subsection{Experiment I: $N_z$ equally spaced zeros on the unit circle}

\begin{figure}[htb]
    \centering
    \includegraphics[width=0.8\textwidth]{ 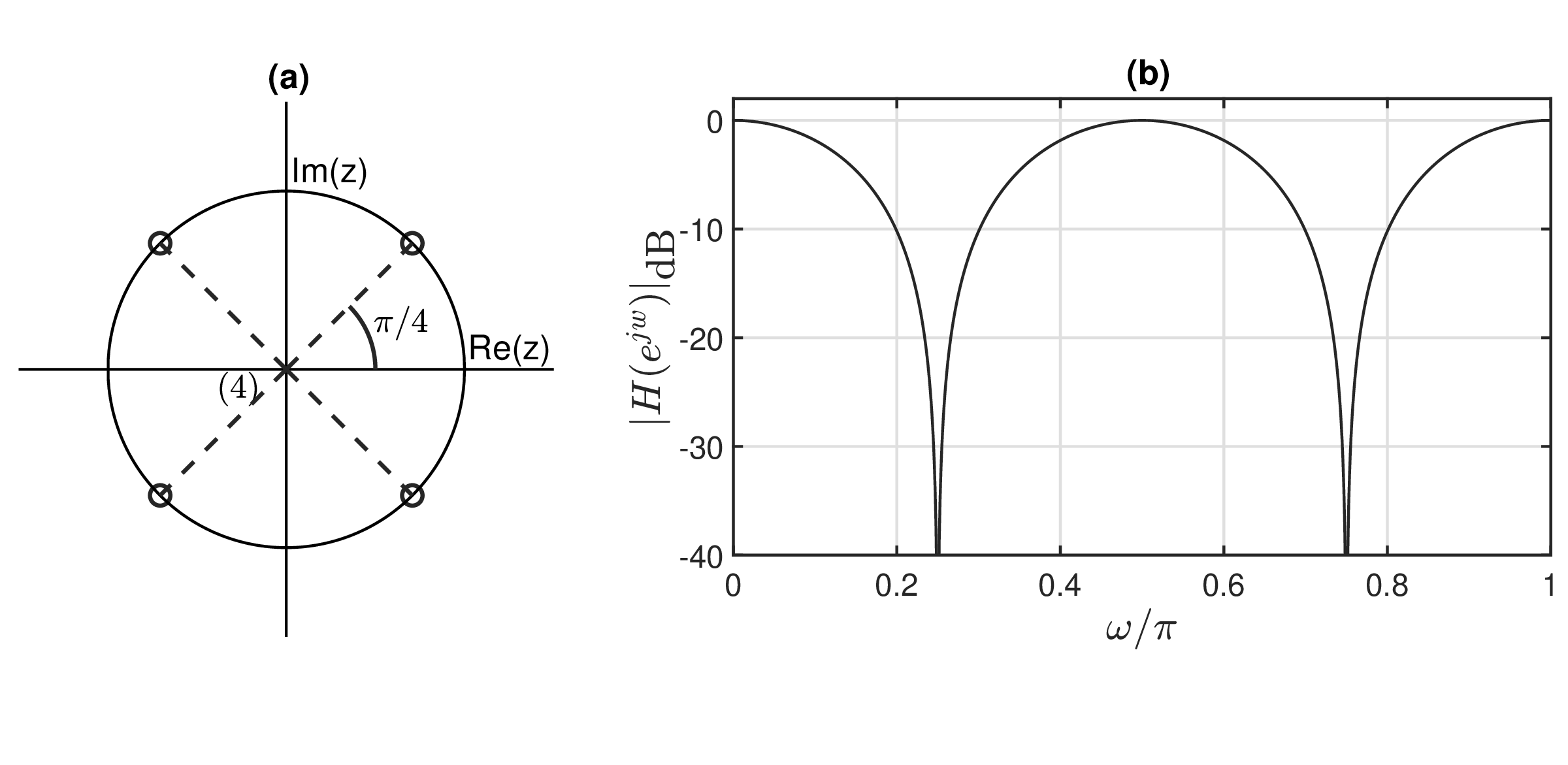}
    \caption{Experiment I - $N_z$ equally spaced zeros on the unit circle. Example for $N_z=4$ and $G=1$: (a) pole-zero plot and (b) magnitude of the frequency response in decibels.}
    \label{fig:zplane1}
\end{figure}

In this first experiment, we consider a filter with equally spaced zeros in the unit circle. In this case, the $N_z$ zeros are spaced by $2\pi/N_z$ and given by \cite{Oppenheim2009}
\begin{equation}
z_k=e^{j\frac{(2k-1)\pi}{N_{z}}},\;\;\;\; k=1,\ldots, N_{z}.
\label{eq:FT_FIR1}
\end{equation}
Note that with this choice, for an even $N_z$ we have $N_z/2$ complex conjugated pairs of zeros and for an odd $N_z$ we have $(N_z-1)/2$ complex conjugated pairs of zeros and a zero at $z=-1$. 
As an example, the zeros for $N_z=4$ are shown in Figure \ref{fig:zplane1}a), and the magnitude of the corresponding frequency response $\left|H\left(e^{j\omega}\right)\right|$ for $G=1$ is shown in Figure \ref{fig:zplane1}b). It has spectral nulls at the frequencies where the zeros are located. 

\begin{figure}[H]
  \includegraphics[width=\textwidth]{ 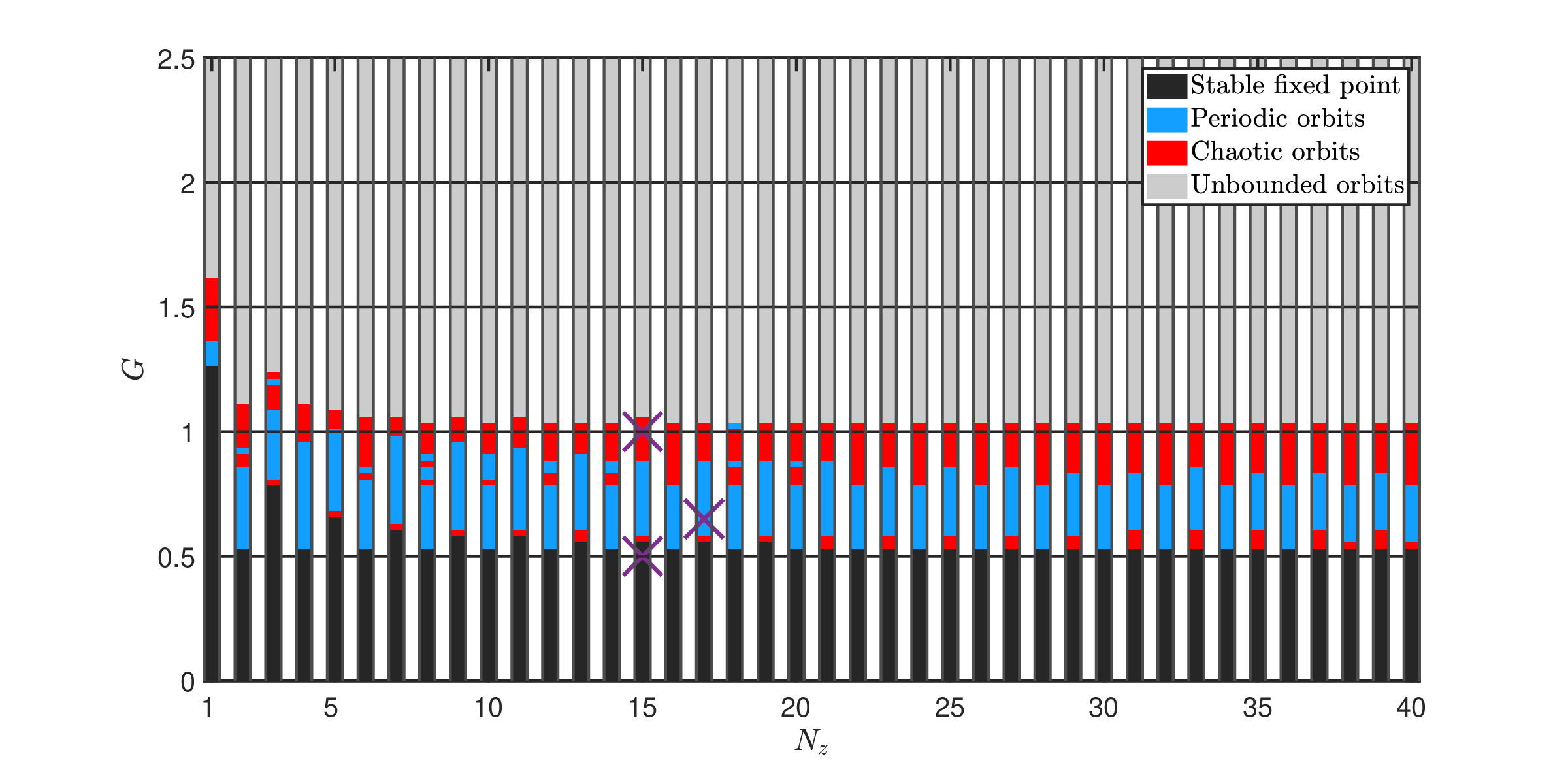}
    \caption{Experiment I - Orbit classification as a function of the filter gain $G$ and the number of zeros $N_z$. Typical orbits for the parameters marked with $\color{purple}\times$ are shown in Figure \ref{fig:teste1_tempo}.}
    \label{fig:teste1}
\end{figure}

The transfer function of such filters is given by
\begin{equation}
H_1(z)=
\begin{cases}
G\displaystyle\prod_{k=1}^{N_z/2}{\frac{z^2-2\cos\left(\frac{(2k-1)\pi}{N_z}\right)z+1}{2\left[1-\cos\left(\frac{(2k-1)\pi}{N_z}\right)\right]z^2}},&\text{even $N_z$}\\\\
G\displaystyle\frac{(z+1)}{2z}\prod_{k=1}^{(N_z-1)/2}\frac{z^2-2\cos\left(\frac{(2k-1)\pi}{N_z}\right)z+1}{2\left[1-\cos\left(\frac{(2k-1)\pi}{N_z}\right)\right]z^2},&\text{odd $N_z$}
\end{cases} \label{eq:Exp1filtro}.
\end{equation}
Since $H_1(z)$ is only  a function of $N_z$ and $G$, we can use \eqref{eq:henonfilt} to numerically analyze the properties of the generated orbits as they vary. The results obtained are shown in Figure \ref{fig:teste1}. 

\begin{figure}[H]
    \centering
    \includegraphics[width=\textwidth]{ 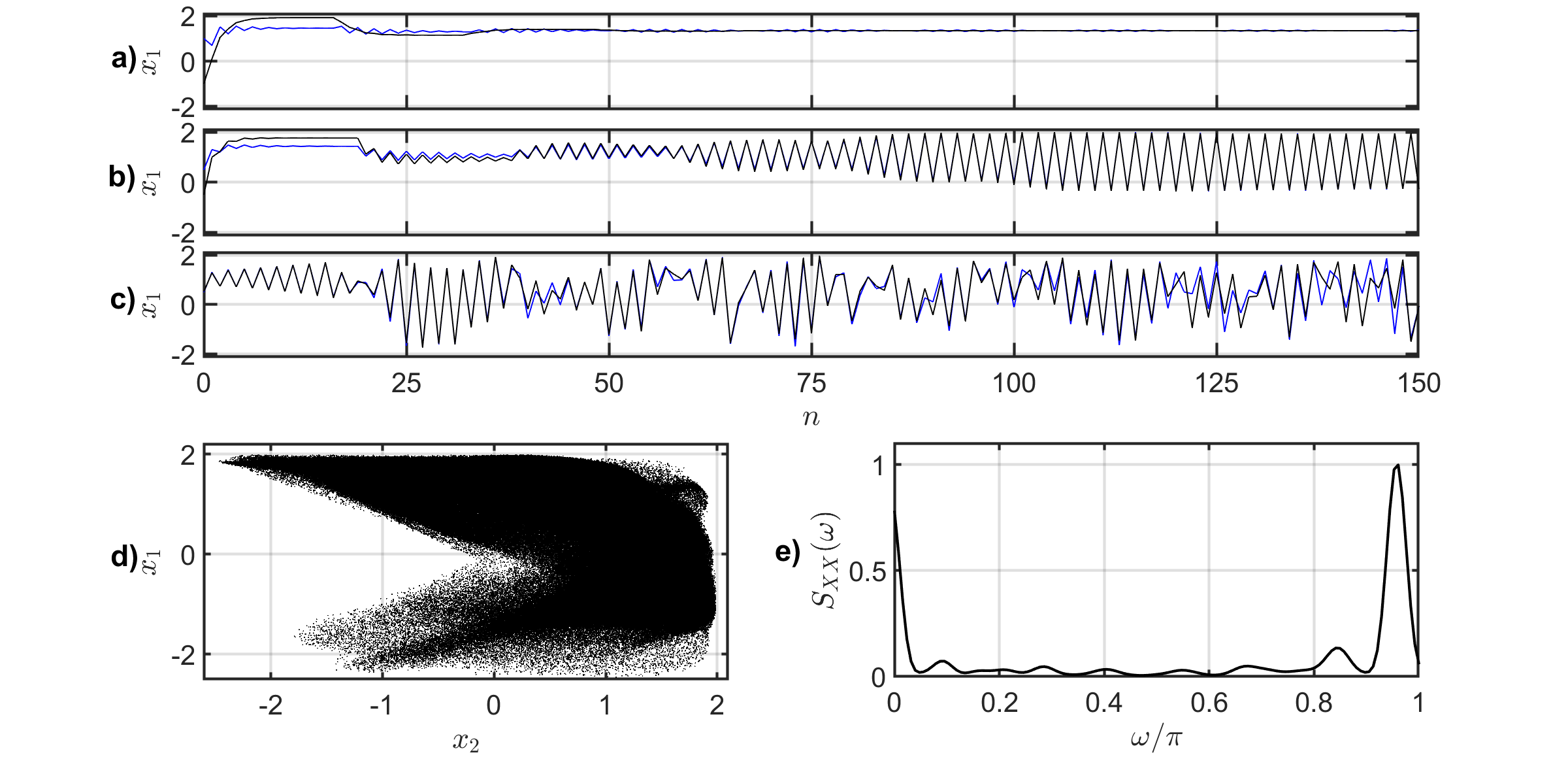}
    \caption{Experiment I - Examples of orbits converging toward a: (a) stable fixed point ($N_{z}=15$, $G=0.5$, and $\mathbf{x}(0)=[1,1,1]$ in black and $\mathbf{x}(0)=[-1,-1,-1]$ in blue); (b) period-2 orbit ($N_{z}=17$, $G=0.65$, and $\mathbf{x}(0)=[0.5,0.5,0.5]$ in black and $\mathbf{x}(0)=[-0.5,-0.5,-0.5]$ in blue); (c) chaotic attractor ($N_{z}=15$, $G=1$, and $\mathbf{x}(0)=[0.5,0.5,0.5]$ in black, and $\mathbf{x}(0)=[0.51,0.51,0.51]$ in blue); \cblue{(d) phase space; and (e) power spectral density for case (c)}. }
    \label{fig:teste1_tempo}
\end{figure}

Typical orbits are exemplified in Figure \ref{fig:teste1_tempo} \cblue{ 
  along with the phase space and power spectral density for the chaotic case. They can be compared with Figures \ref{fig::exemplohenon}(b) and (c). One can clearly see that the insertion of the filter qualitatively changes the phase space, as expected, since the filter is in the feedback loop of the chaotic map generator. This change reflects the increase in the dimensionality of the global system caused by the filter. In this case, the phase space is a 2D projection of a higher dimensional trajectory. The power spectral density also changes significantly. One should note, however, that this cannot be directly interpreted as a filtering of the original map signal by the filter \eqref{eq:Exp1filtro}, since it is in the feedback loop of a nonlinear system.}  
  
\begin{figure}[H]
    \centering
  \includegraphics[width=\textwidth]{ 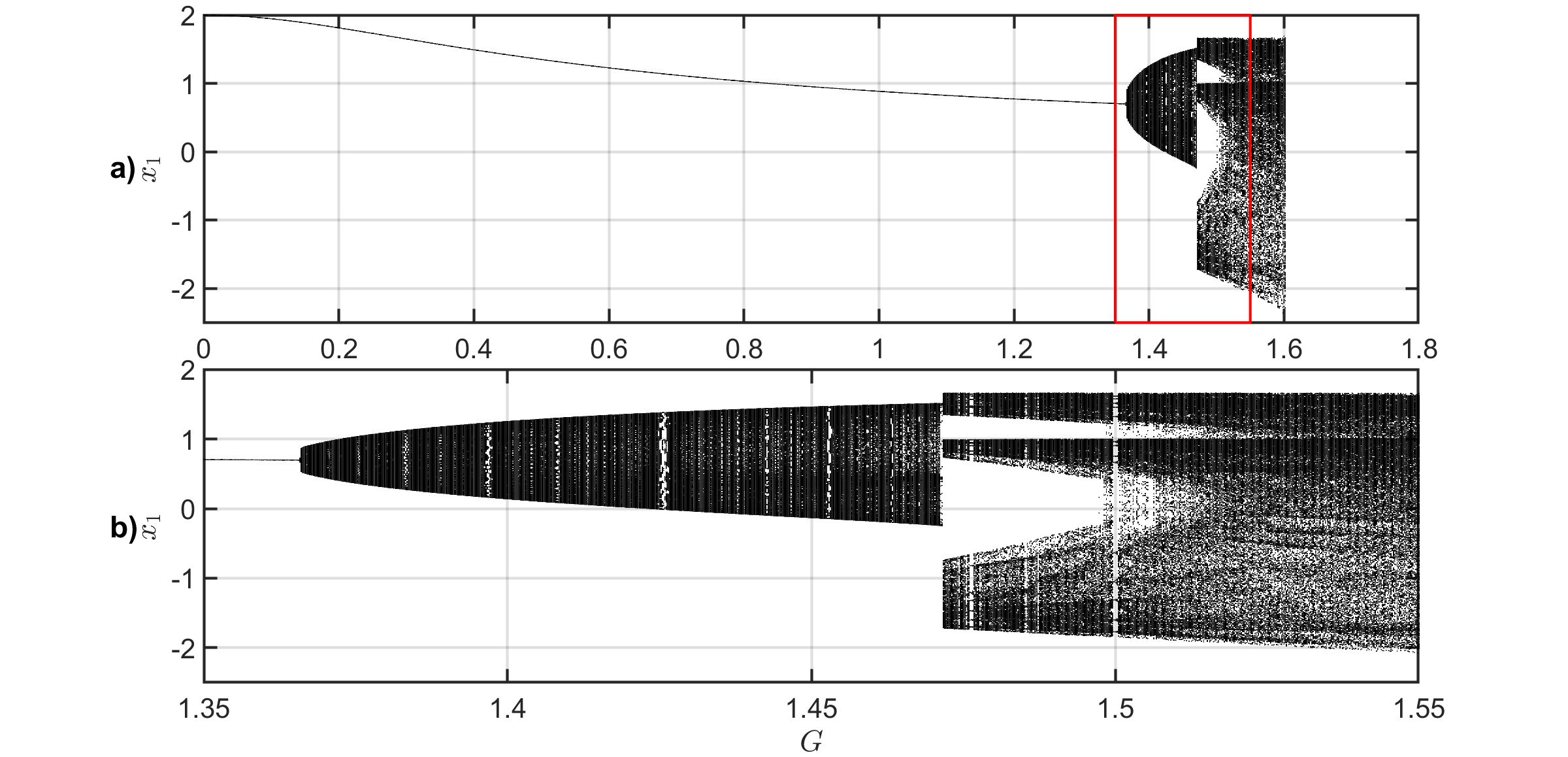}
    \caption{Experiment I - (a) Bifurcation diagram  for $N_z=1$ and \cblue{(b) Zoom on the rectangle 
highlighted in (a).}}
    \label{fig:bifurcação}
\end{figure}

  A bifurcation diagram for  $N_z=1$ is illustrated in Figure~\ref{fig:bifurcação}. \cblue{The strategy of following the attractor was used: for each value of $G$, the initial condition was a point of the attractor obtained from its previous value.}

We can observe in Figure \ref{fig:teste1} that as the gain increases from $G=0$, there is an evolving pattern of orbits, starting with convergence to a stable fixed point, passing through a set of periodic orbits, and arriving at chaotic orbits. Finally, as $G$ continues to increase, the orbits diverge.  This pattern can also be seen for $N_z=1$ in the bifurcation diagram in Figure~\ref{fig:bifurcação}. The width of the range of $G$ values in which the chaotic behavior is present does not change systematically with $N_z$ for $N_z>1$. The value of $G$  has more influence on the asymptotic properties of the orbits than $N_z$. For example, $G=1$ results in chaotic orbits for all $N_z>1$ and for $G<0.5$ the fixed point is stable independently of $N_z$. 

Another interesting aspect of Figure \ref{fig:teste1} is that, in general, for odd $N_z>1$ when $\mathbf{p}^+$ becomes unstable, a window of chaotic orbits appears immediately. For even $N_z$, the loss of stability of the fixed point is followed by periodic orbits.

Finally,  note that for $N_z=1$  the only zero is $z_1=-1$. Thus, from \eqref{eq:fat2}, \begin{equation}
    H_1(z)=G\frac{z+1}{2z}
\label{eq:H1Nz1}\end{equation} and the last equation in \eqref{eq:henonfilt} is $x_3(n+1)=c_0x_1(n+1)+c_1x_1(n)$ with $c_0=c_1=G/2$. Thus, the dynamics of this particular case can be directly compared to Figures~1 and 2 of \cite{borges2022f}. As expected, our sequence of stable fixed-point, periodic orbits, chaotic orbits, and divergence agrees with that earlier work. 

\subsection{Experiment II: a pair of complex conjugate zeros on the unit circle}

\begin{figure}[htb]
    \centering
\includegraphics[width=0.8\textwidth]{ 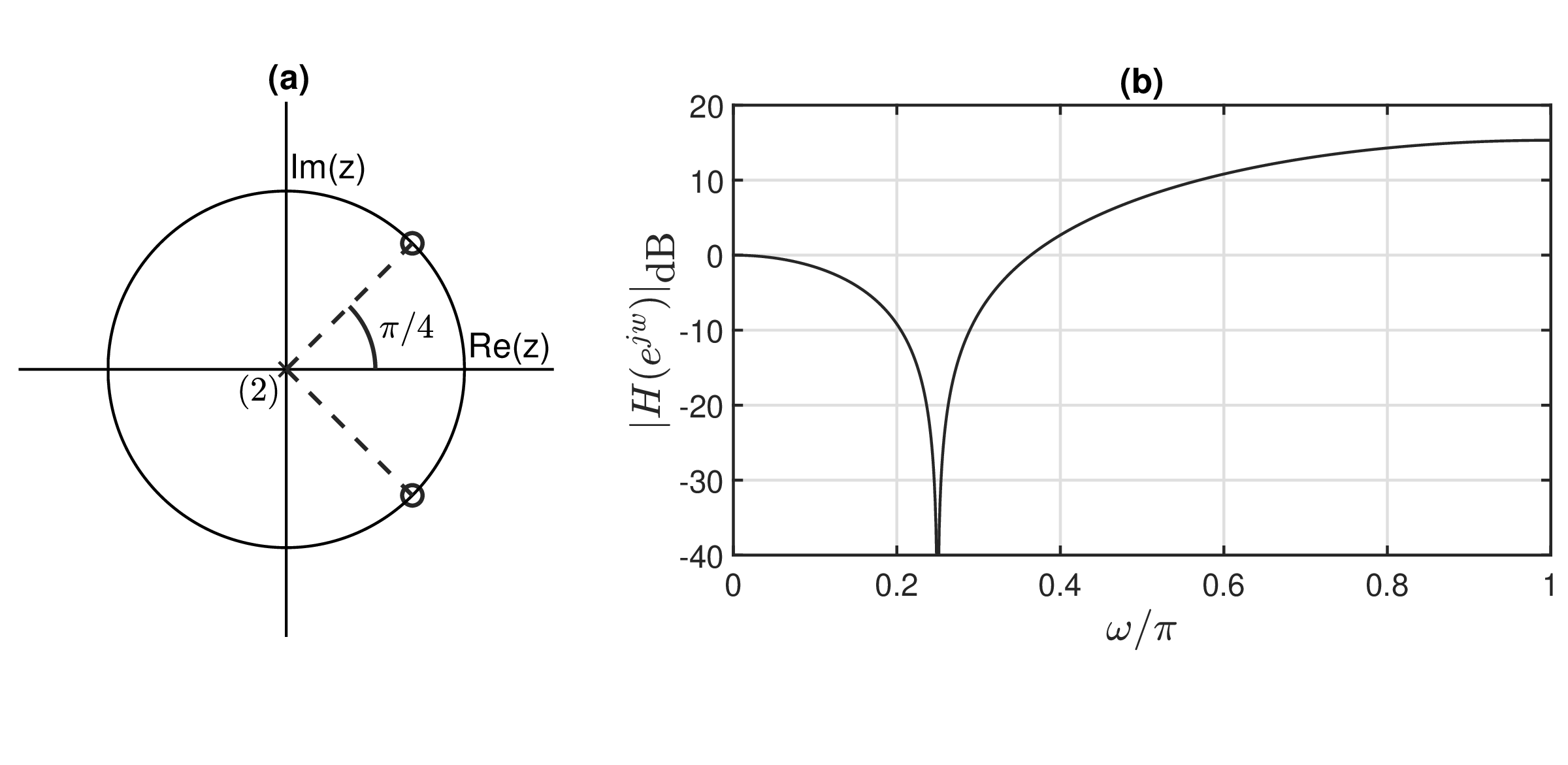}
    \caption{Experiment II - A pair of complex conjugate zeros on the unit circle. Example for $\omega_0=\pi/4$ and $G=1$: (a) pole-zero plot and (b) magnitude of the frequency response in decibels.}
    \label{fig:zplane3}
\end{figure}

In this experiment, we consider only $N_z=2$ zeros in $z_{1,2}=e^{\pm j\omega_0}$
with $0<\omega_0\leq\pi$. Thus, from \eqref{eq:fat2},
\begin{equation}
H_2(z)=G\frac{\left(z^2-2\cos\omega_0z+1\right)}{2z\left(1-\cos\omega_0\right)}\label{eq:filtroEx2}.
\end{equation}
Figure \ref{fig:zplane3} shows the pole-zero plot and the magnitude of the frequency response of \eqref{eq:filtroEx2} for $\omega_0=\pi/4$ and $G=1$. This filter is known in the literature as a notch filter, since its frequency response has only one spectral null at $\omega=\omega_0$ \cite{Oppenheim2009}.

\begin{figure}[H]
    \centering
   \includegraphics[width=\textwidth]{ 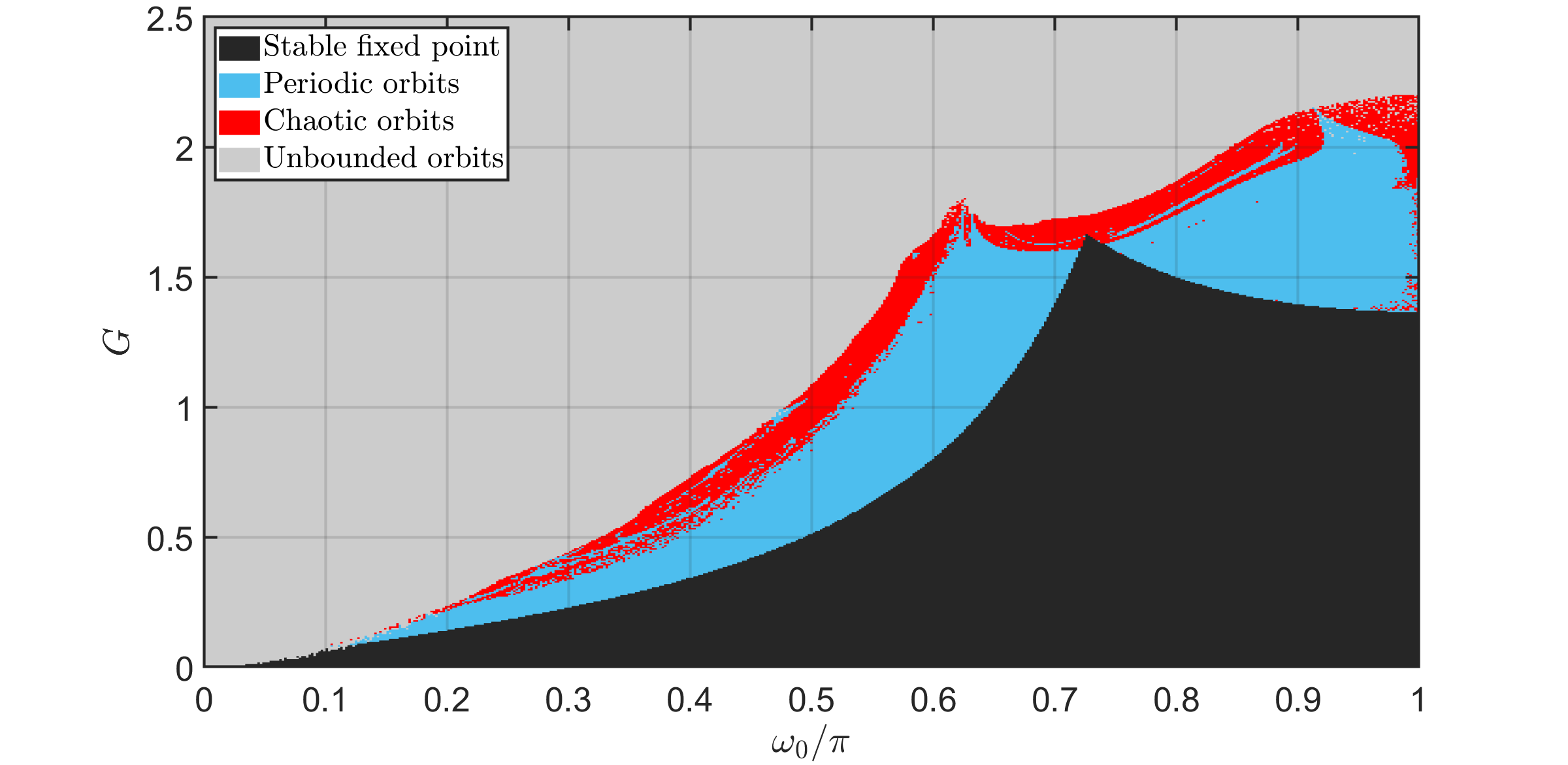}
    \caption{
    Experiment II - Classification of
    each parameter set according to the orbit 
    as a function of ${\omega_0}/{\pi}$ and $G$. 
    }
    \label{fig:teste2}
\end{figure}

Varying the frequency $\omega_0$ and the gain $G$, we compute the largest Lyapunov exponent for each set of parameters, allowing us to classify it according to the orbits, as shown
in Figure~\ref{fig:teste2}. 
We can observe in this figure that for higher $\omega_0$, there is a larger set of $G$ values that results in a stable fixed point. On the other hand, unbounded orbits occur more often for lower $\omega_0$. The region of periodic orbits is larger than the region of chaotic orbits that appears between the periodic and divergent orbits for $\omega_0>0.2\pi$. 

\begin{figure}[H]
    \centering
   \includegraphics[width=\textwidth]{ 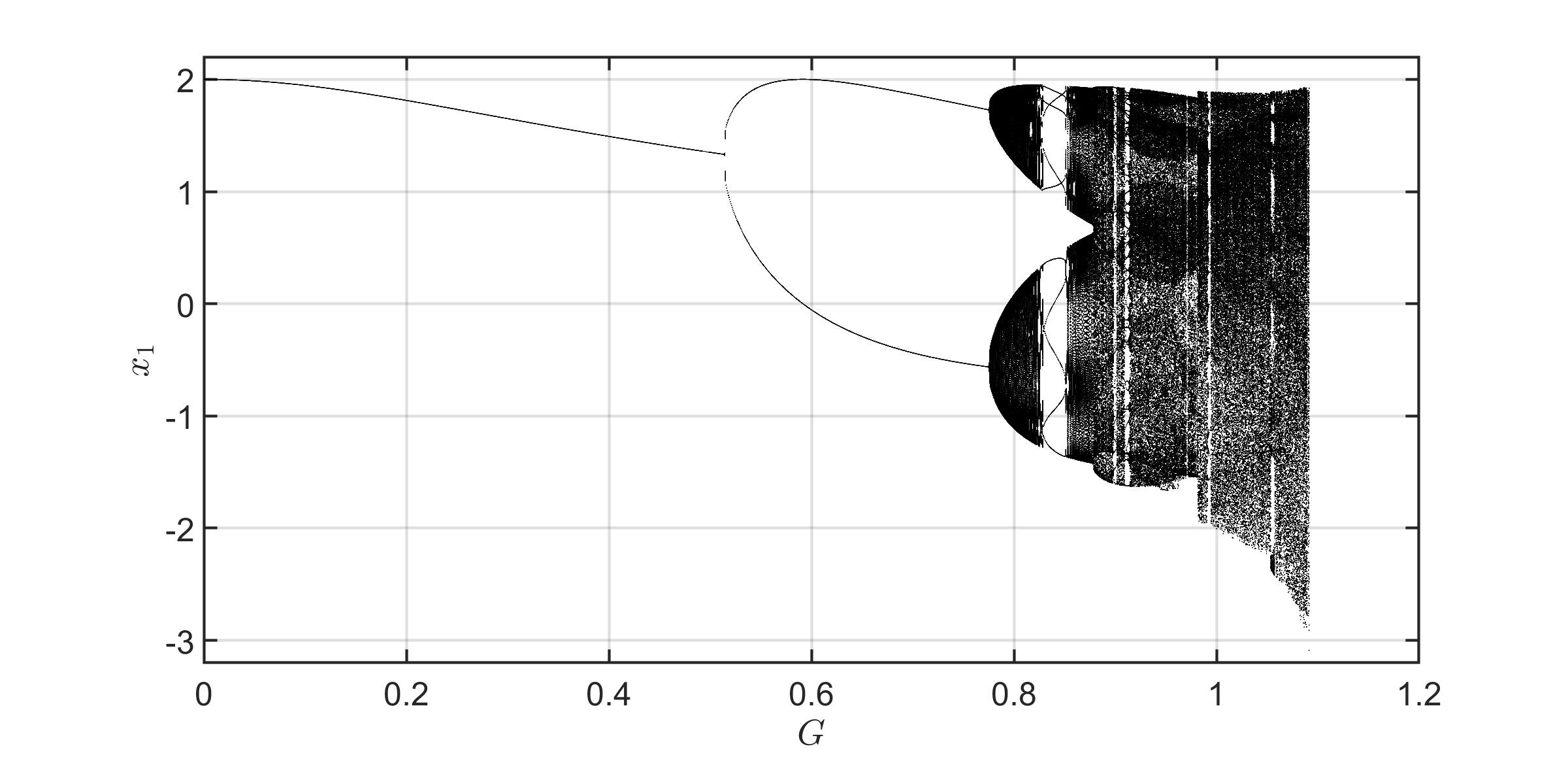}
    \caption{Experiment II - Bifurcation diagram for $\omega_0/\pi = 0.5$.
    }
    \label{fig:teste2_bifurcacao}
\end{figure}

The bifurcation diagram as a function of $G$  for $\omega_0=\pi/2$ is shown in Figure~\ref{fig:teste2_bifurcacao}. \cblue{As in Figure \ref{fig:bifurcação}, we employed the strategy of following the attractor.} We can observe that the system has a stable fixed point for $G\in [0.00, 0.51]$ and a stable period-2 orbit for $G\in [0.51, 0.77]$. For $0.77<G<1.10$ the diagram presents chaotic orbits with periodic windows. For $G\geq 1.10$ the system has unbounded orbits. We should note that the transfer function $H_2(z)$
 for $\omega_0=\pi/2$ is the same as $H_1(z)$ of \eqref{eq:Exp1filtro} with $N_z=2$.


\subsection{Experiment III: a pair of complex conjugate zeros on the unit circle and a zero at $z=-1$\label{subsec:expIII}}

\begin{figure}[htb]
    \centering
    \includegraphics[width=0.8\textwidth]{ 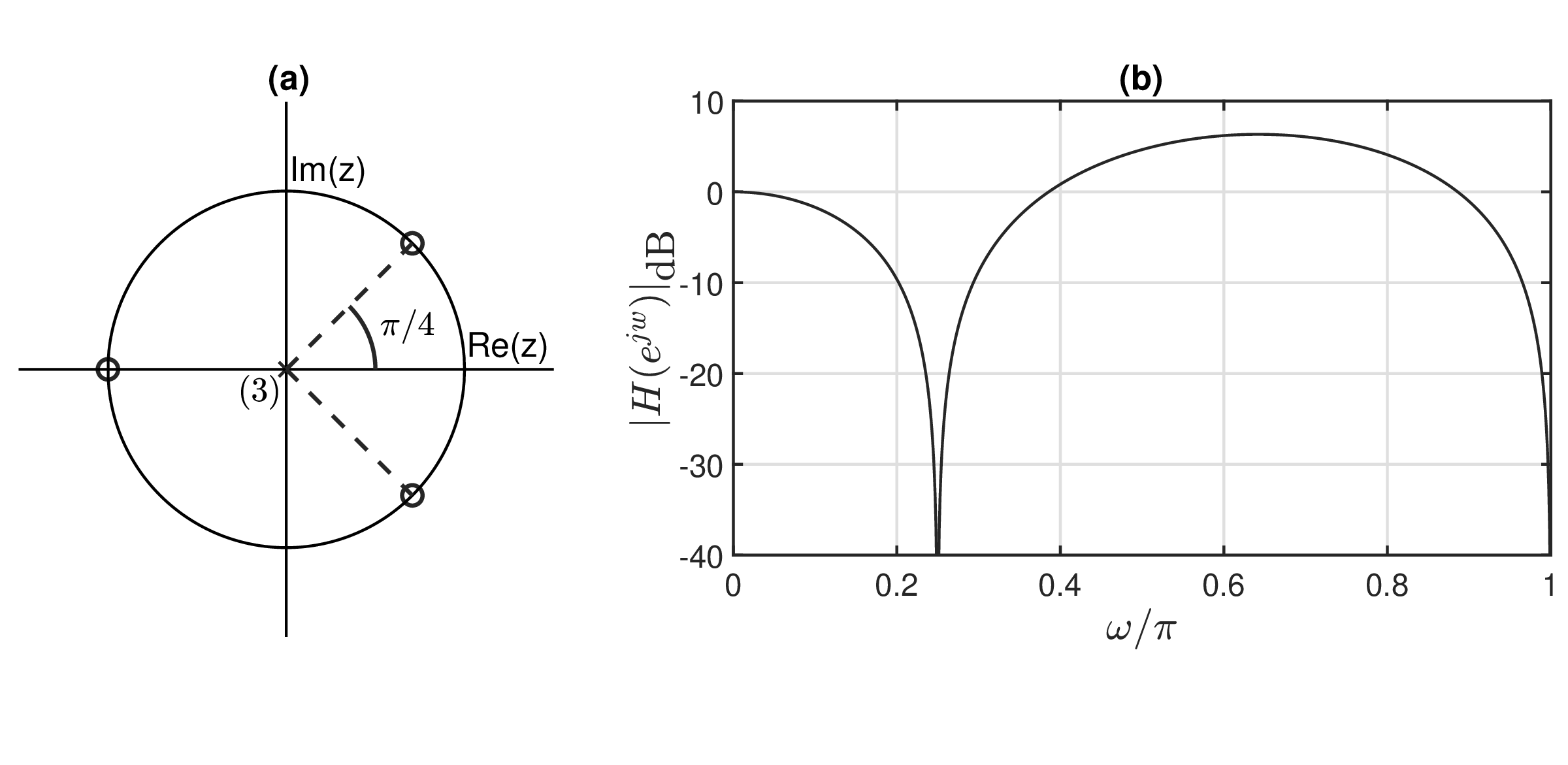}
    \caption{Experiment III - A pair of complex conjugate zeros at different frequencies on the unit circle and a zero at $z=-1$. Example for $\omega_0=\pi/4$ and $G=1$: (a) pole-zero plot and (b) magnitude of the frequency response in decibels.}
    \label{fig:zplane_exp_3}
\end{figure}

In this experiment, we have $N_z=3$ zeros, which are $z_{1,2}=e^{\pm j\omega_0}$, $0<\omega_0\leq\pi$, and $z_3=-1$. Thus, replacing these zeros in \eqref{eq:fat2}, we obtain the  transfer function
\begin{equation}
    H_3(z)=G\frac{\left(z^2-2\cos\omega_0z+1\right)(z+1)}{4z\left(1-\cos\omega_0\right)}.\label{eq:filtroEx3}
\end{equation}
Figure \ref{fig:zplane_exp_3} shows the pole-zero plane and the magnitude of the  frequency response of \eqref{eq:filtroEx3} for $\omega=\pi/4$ and $G=1$.

This experiment is similar to Experiment II with the notable difference that we have an odd $N_z$. As we noted in Experiment I, the parity of $N_z$ influences the dynamics (Figure \ref{fig:teste1}). 
The frequency response has two spectral nulls at $\omega=\omega_0$ and at $\omega=\pi$.    

\begin{figure}[htb]
    \centering
    \includegraphics[width=\textwidth]{ 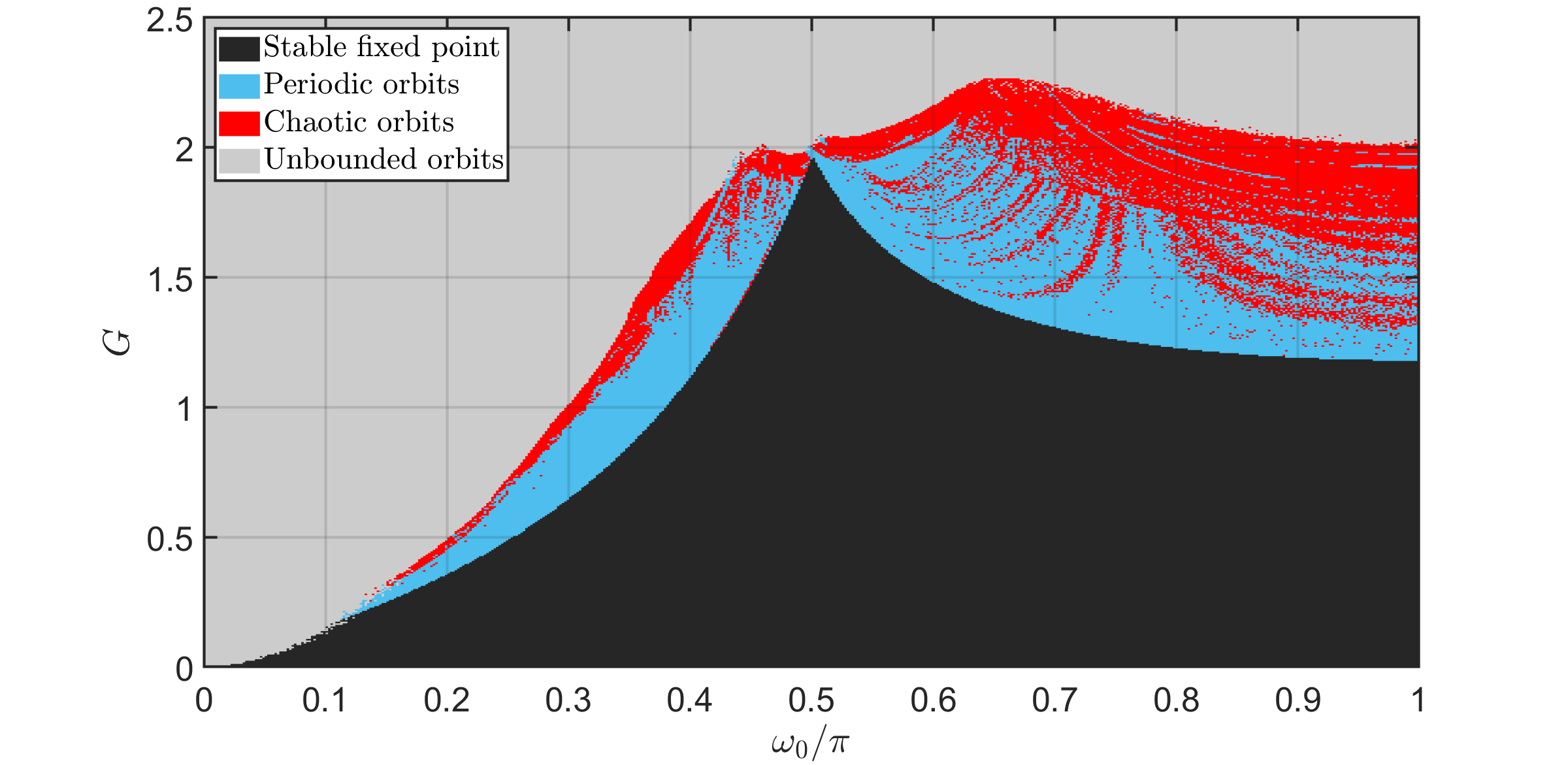}
    \caption{Experiment III - Classification of
    each parameter set according to the orbit
    as a function of ${\omega_0}/{\pi}$ and $G$.}
    \label{fig:teste3}
\end{figure}

Again, we can study the generated signals as a function of $\omega_0$ and $G$. The results are presented in Figure~\ref{fig:teste3}. Comparing Figures~\ref{fig:teste2}~and~\ref{fig:teste3}, we can see an increase in the chaotic region with an advance of the chaotic orbit regions over the regions where periodic orbits predominate, which was not the case in Experiment~II. 

\subsection{Experiment IV: $N_z$ zeros at $z=-1$}

\begin{figure}[htb]
    \centering
    \includegraphics[width=0.8\textwidth]{ 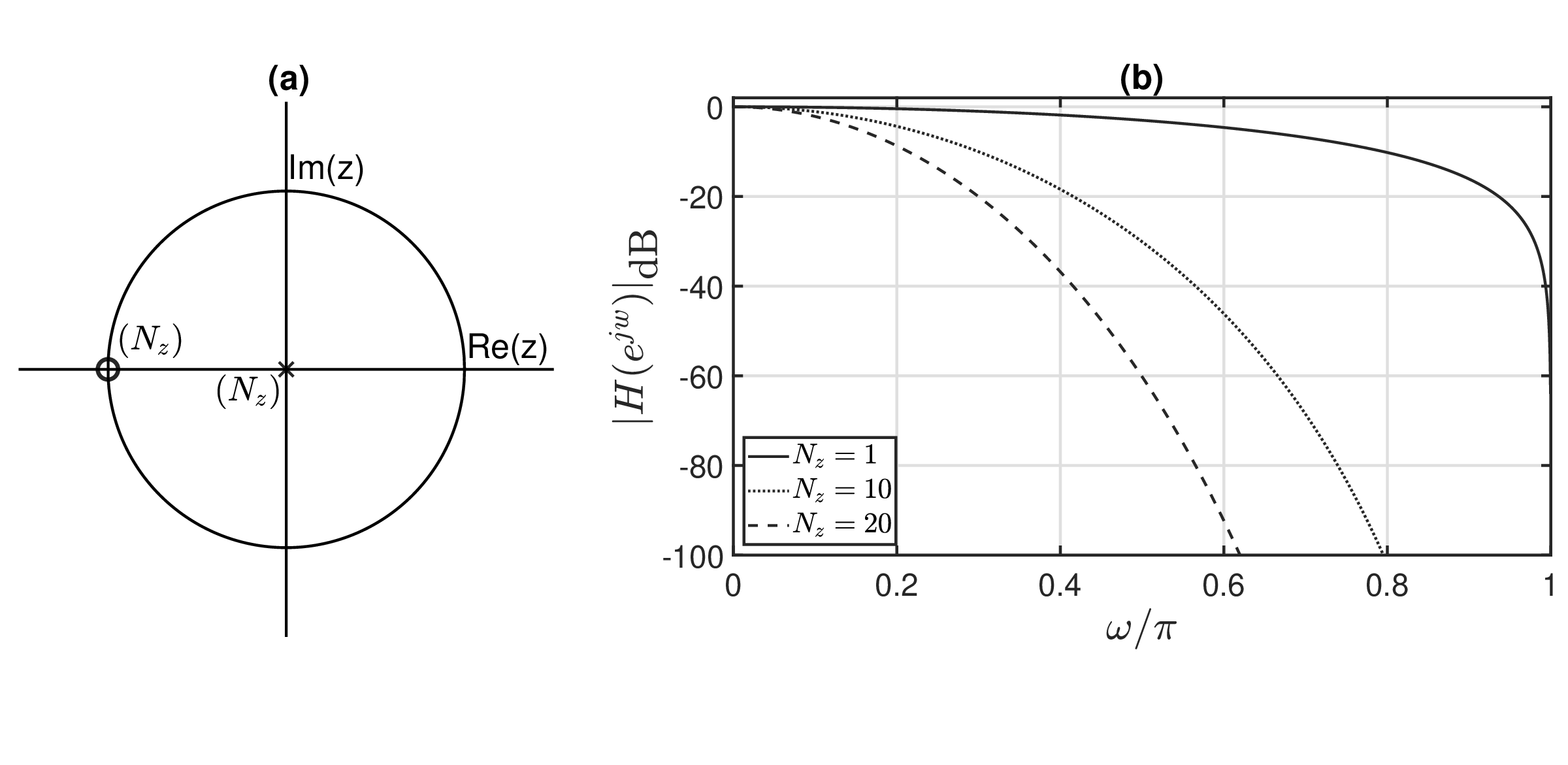}
    \caption{Experiment IV - $N_z$ zeros at $z=-1$. (a) Pole-zero plot and (b) Magnitude of frequency response. Examples with $N_z=1$, $N_z=10$ e $N_z=20$ and $G=1$.}
    \label{fig:zplane4}
\end{figure}

In the fourth experiment, we analyze the influence of the number of zeros in $z=-1$. In this case, the transfer function of the FIR filter assumes the form
\begin{equation}
    H_4(z)=\frac{G}{2^{N_z}}\frac{(z+1)^{N_z}}{z^{N_z}}.\label{eq:H4}
\end{equation}
Figure~\ref{fig:zplane4} shows the pole-zero plot and the magnitude of the frequency response of \eqref{eq:H4} for $N_z=1$, $N_z=10$, and $N_z=20$ and $G=1$. We can observe that the larger $N_z$, the faster the magnitude of the frequency response decays to zero at $\omega=\pi$. 

\begin{figure}[htb]
    \centering
    \includegraphics[width=\textwidth]{ 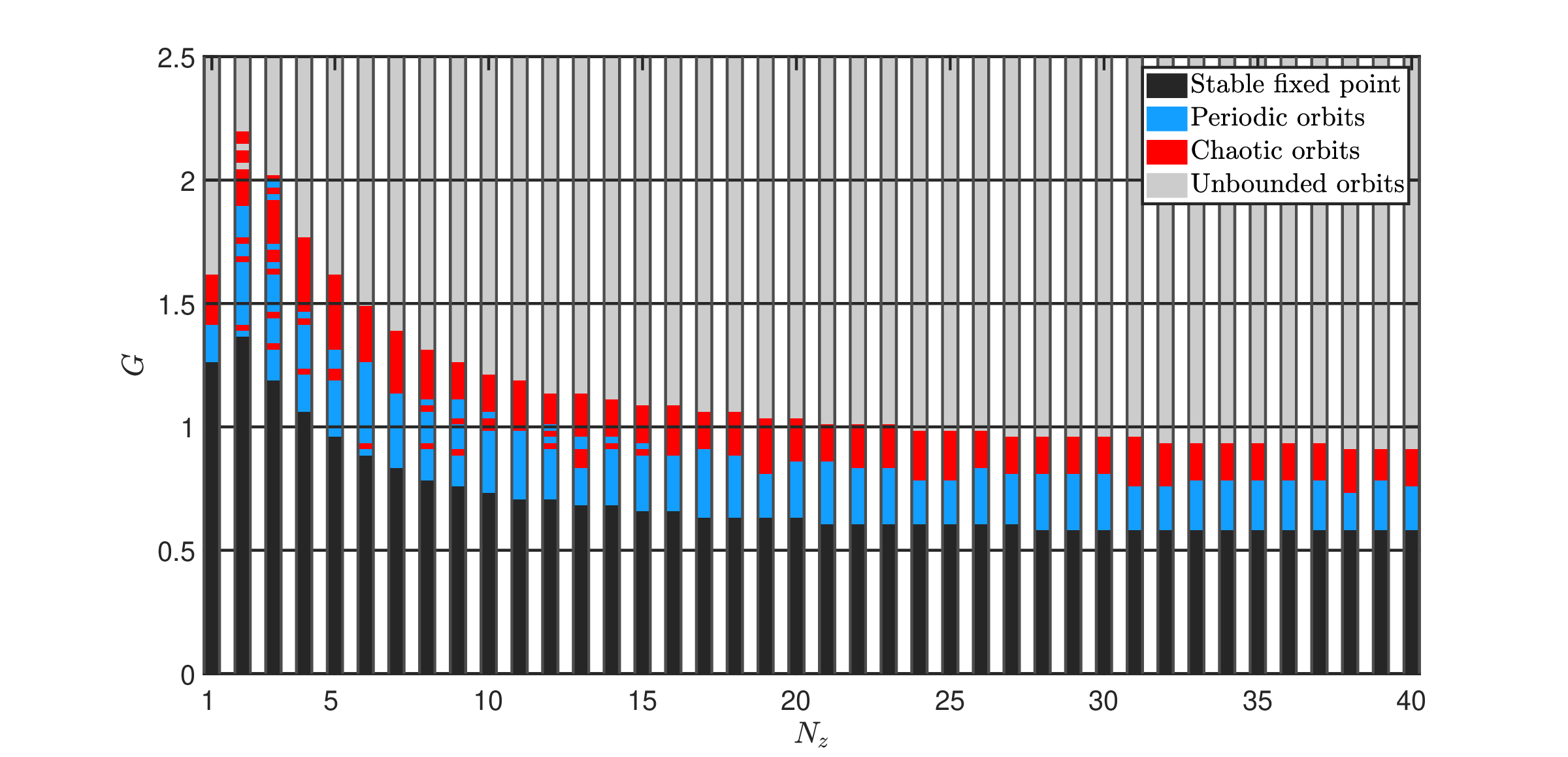}
    \caption{Experiment IV - Classification of
    each parameter set according to the orbit
    as a function of $N_{z}$ and $G$.}
    \label{fig:teste4}
\end{figure}

In Figure \ref{fig:teste4}, it can be seen that for $N_z\leq3$ there is a growth of the region where there is a stable fixed point, however, as $N_z$ increases the region with divergent orbits gets larger, and the chaotic and periodic regions are limited to a gain range between $0.5$ and $1$. In the case of $G\leq0.5$, there is a guarantee of finding stable fixed points, as in Experiment I, Figure \ref{fig:teste1}. Note that the case $N_z=1$ is the same for Experiments I and IV.

\subsection{Experiment V: Low-pass filter designed using \cblue{the} Hamming window}

\begin{figure}[htb]
    \centering
    \includegraphics[width=0.8\textwidth]{ 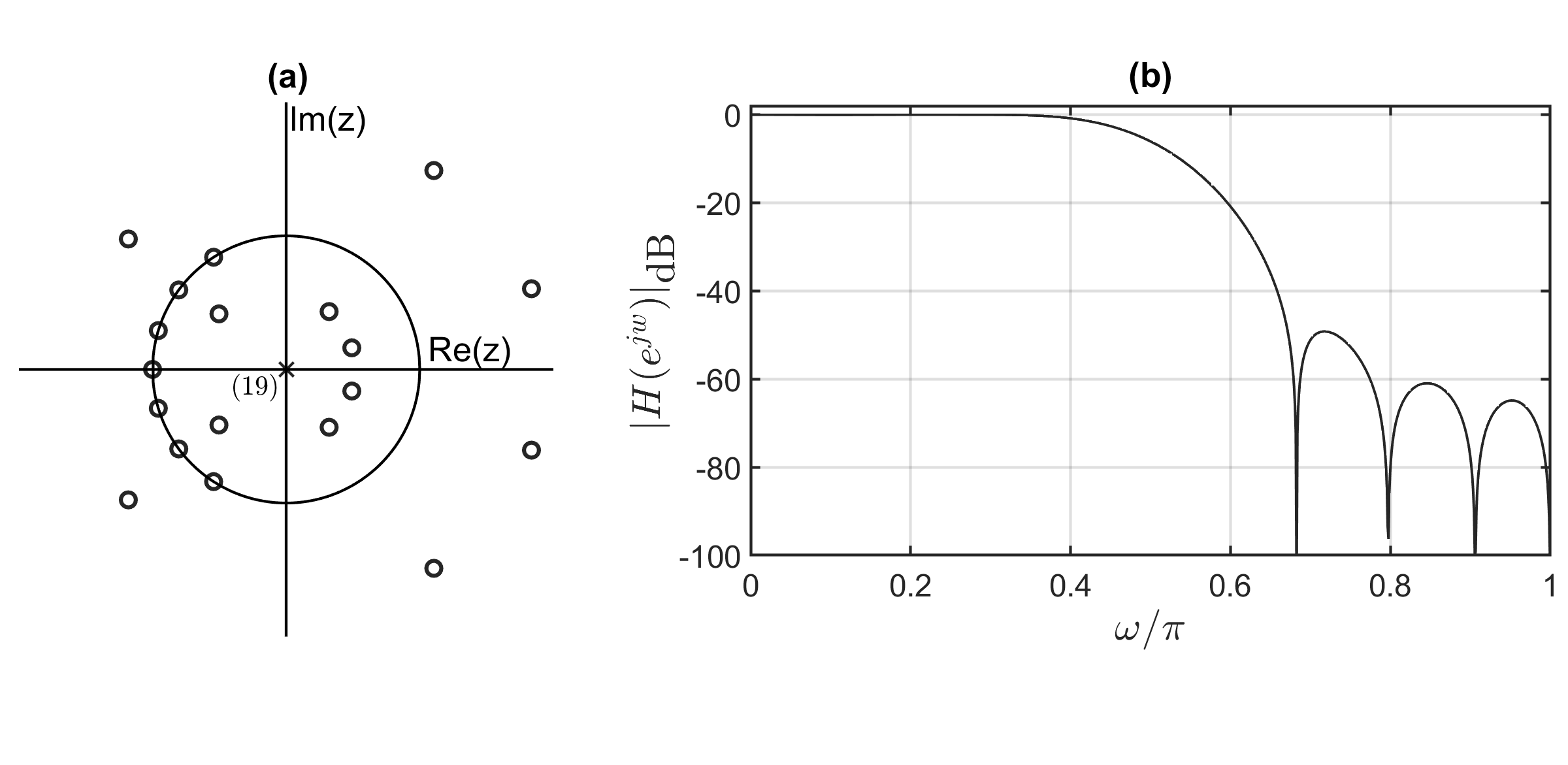}
    \caption{Experiment V - (a) Pole-zero plot and (b) Magnitude of frequency response. Example with $N_z=19$, $\omega_c=0.5\pi$, and $G=1$.}
    \label{fig:zplane5}
\end{figure}

To deal with more practical filters,  in this section, we consider lowpass FIR filters designed with different numbers of zeros, cutoff frequency $\omega_c$, $0<\omega_c<\pi$, $G=1$, and using the Hamming window \cite{Oppenheim2009}. The transfer function of such filters is given by \cite{Oppenheim2009}
\begin{equation}
H_5(z)=\frac{1}{z^{N_z}}\displaystyle\frac{\displaystyle\sum_{j=0}^{N_z}{\rm sinc}\,[\omega_c(j-N_z/2)]\cdot\left[0.54-0.46\cos\left(2\pi j/N_z\right)\right] z^{N_z-j}}
{\displaystyle\sum_{j=0}^{N_z}{\rm sinc}\,[\omega_c(j-N_z/2)]\cdot\left[0.54-0.46\cos\left(2\pi j/N_z\right)\right]},\label{eq:H5}
\end{equation}
where ${\rm sinc}(x)\triangleq\sin(x)/x.$
Figure \ref{fig:zplane5} 
shows the pole-zero plot and the magnitude of the frequency response of \eqref{eq:H5} for $N_z=19$ and $\omega_c=0.5\pi$. We can observe that the zeros of the filter that are not \cblue{in} the unit circle occur in conjugate reciprocal pairs, which is a necessary condition for obtaining linear-phase FIR filters \cite{Oppenheim2009}. \cblue{Regarding} the magnitude of the frequency response, this example corresponds to a gain in the passband between $0$~ dB and $0.054$~ dB and a stopband attenuation of at least 
$-49$~dB~\cite{Oppenheim2009}.

\begin{figure}[htb]
    \centering
    \includegraphics[width=\textwidth]{ 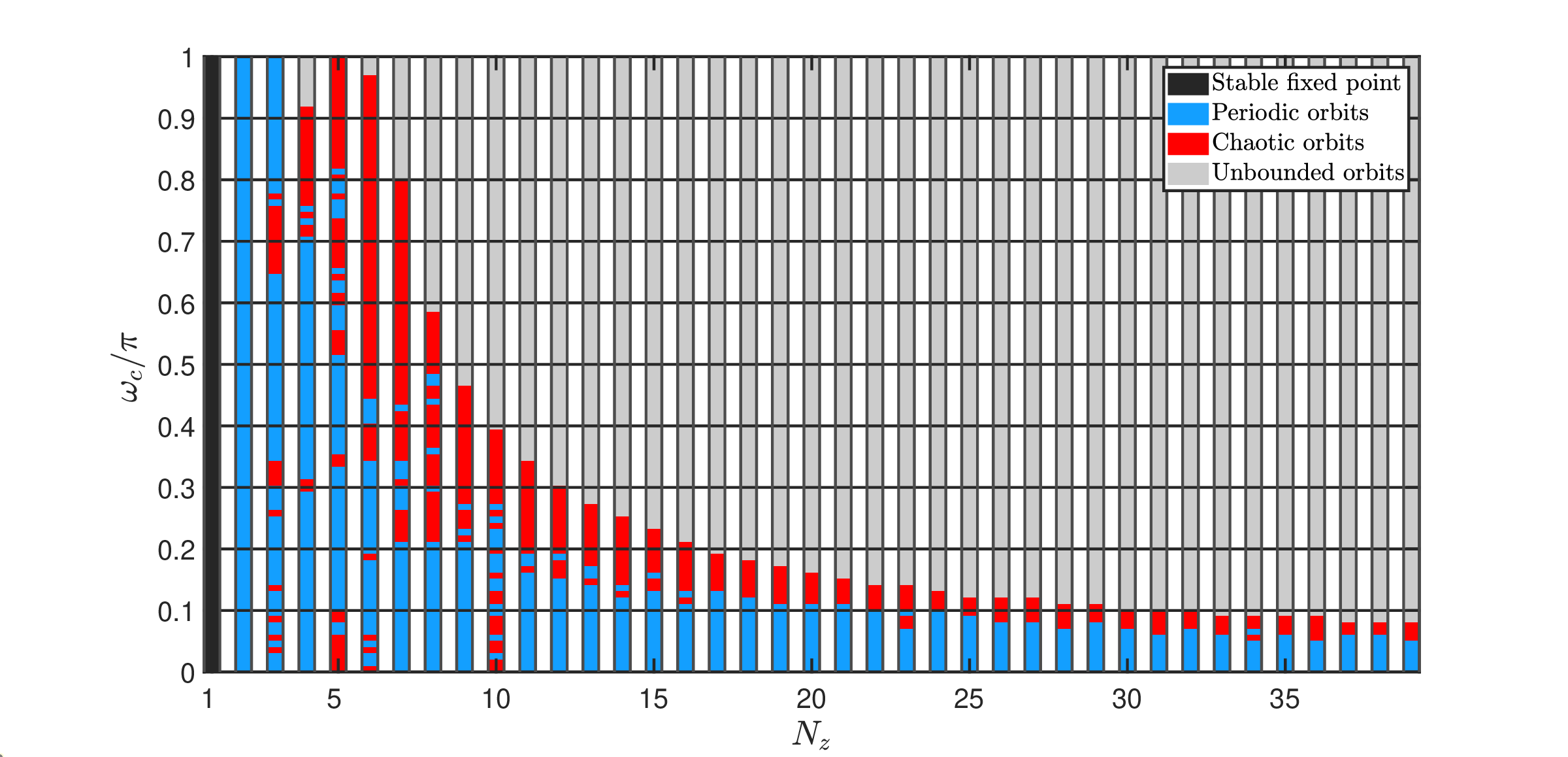}
    \caption{Experiment V - Classification of
    each parameter set according to the orbit
    as a function of $N_{z}$ and $\omega_c$.}
    \label{fig:teste5_hamming}
\end{figure}

Varying $\omega_c$ and $N_z$, we again calculate the largest Lyapunov exponent for each set of parameters, leading to the orbit classification shown in Figure~\ref{fig:teste5_hamming}. For $N_z=1$, $H_5(z)$ does not depend on $\omega_c$.  Thus, the whole case $N_z=1$ in Figure \ref{fig:teste5_hamming} corresponds in fact to the single filter \eqref{eq:H1Nz1} with $G=1$. The stable fixed point agrees with the result for $N_z=1$ and $G=1$ in Figure \ref{fig:teste1}. For $N_z=2$, we observe the convergence to a periodic orbit over the entire range of $\omega_c$. 
As $N_z$ increases, the range of parameters that leads to periodic orbits decreases, and these orbits appear for low $\omega_c$.
Chaotic regions appear for $N_z\geq3$ and this is the most observed case for $N_z=6$ and $N_z=7$. For $8\leq N_z \leq 40$, the convergence to chaotic orbits is less frequent while divergent orbits predominate for $N_z\geq 8$.


\section{Conclusions}
\label{section_conclusion}

In recent years, several applications of dynamic systems that generate chaotic signals have appeared in engineering and applied physics. In some of them, limiting the frequency content of chaotic signals is fundamental, which can be achieved by inserting linear filters in their description. However, the influence of these filters on the asymptotic behavior of the trajectories is not fully understood. In this sense, in this work, we present numerical experiments
using the recently proposed filtered Hénon map \cite{borges2022f} to evaluate which characteristics (gain, position, parity, and number of zeros) of an FIR filter are relevant in terms of asymptotic dynamics.

Interestingly, we found that the fixed points of the filtered Hénon map using a generic FIR filter depend only on the filter coefficients through the filter gain $G$. Specific values of $c_j$ do not affect them. However, their stability is more subtle and depends on the composition of the filter coefficients. Our numerical experiments show that for the different filters tested, there is a range of $G$ that produces chaotic and periodic orbits. Above this range, the orbits diverge and below it, the fixed points are stable.

In summary, the incorporation of linear filters into a dynamical system can significantly change the nature of the generated signals. \cblue{When using maps that produce signals known to be chaotic, it cannot be assumed that this property will be maintained when filters, even linear ones, are introduced in the system. This variation in dynamics as a function of the filters employed warns of the need to carefully evaluate proposals for bandlimited chaos-based communication systems.} 

\cblue{This article leaves several possibilities of future research. For instance: i) 
the effects of the introduction of filters on
the properties associated with the generation of pseudorandom sequences using the Hénon map \cite{MeranzaCastillon2019, Kocarev2003}; ii) the possibility of using a symmetric version of the Hénon map \cite{Butusov2018a, tutueva2022fast}; iii) the influence of digital noise and truncation on the chaotic properties of the orbits; and iv) the effects of the insertion of infinite impulse response (IIR)  or nonlinear filters into the chaos generator.}

\appendix
\section{Jacobian matrices}
\label{Apen:Jacobian}

In this appendix, we obtain the Jacobian used to perform the stability calculations of the numerical experiments of Section~\ref{section_experiments}. 

For $N_z>3$, we define the following
\begin{equation}
\left\{\begin{aligned}
x_{4}(n+1) & \triangleq x_{2}(n) \\
x_{5}(n+1) & \triangleq x_{4}(n) \\
x_{6}(n+1) & \triangleq x_{5}(n) \\
& \vdots \\
x_{N_{z}+1}(n+1) & \triangleq x_{N_{z}}(n)
\end{aligned}\right..
\label{eq:var_aux} 
\end{equation}
Then, replacing $x_{1}(n+1)$ into the last equation of \eqref{eq:henonfilt} and using \eqref{eq:var_aux}, we rewrite the system as
\begin{equation}
\left\{\begin{aligned}
x_{1}(n+1)=& \alpha-(x_{3}(n))^{2}+\beta x_{2}(n) \\
x_{2}(n+1)=& x_{1}(n) \\
x_{3}(n+1)=& c_{0}\left[\alpha-(x_{3}(n))^{2}+\beta x_{2}(n)\right]+c_{1}x_{1}(n)\\
&+c_{2}x_{2}(n)+c_{3}x_{4}(n)+c_{4}x_{5}(n)\\
&+\cdots+c_{N_{z}} x_{N_{z}+1}(n) \\
& \vdots \\
x_{4}(n+1)=& x_{2}(n) \\
x_{5}(n+1)=& x_{4}(n) \\
x_{6}(n+1)=& x_{5}(n) \\
& \vdots \\
x_{N_{z}+1}(n+1)=& x_{N_{z}}(n)
\end{aligned}\right..
\label{eq:henonfiltrated}
\end{equation}

The Jacobian matrix  allows the calculation of the Lyapunov exponents of the bandlimited Hénon map~\cite{Alligood2000}. For the system described in \eqref{eq:henonfiltrated} the Jacobian is given by
\begin{equation}
\mathbf{J}_{N_z}\left(\mathbf{x}(n)\right)=
\begin{bmatrix}
0 & \beta & -2x_{3}(n) & 0 & \cdots & 0 & 0 \\
1 & 0 & 0 & 0 & \ldots & 0 & 0 \\
c_{1} & c_{0} \beta+c_{2} & -2c_{0} x_{3}(n) & c_{3} & \ldots & c_{N_z-1} & c_{N_z} \\
0 & 1 & 0 & 0 & \ldots & 0 & 0 \\
0 & 0 & 0 & 1 & \ldots & 0 & 0 \\
& & \vdots & & \ddots &  & \vdots\\
0 & 0 & 0 & 0 & \ldots & 1 & 0 
\end{bmatrix}_{\left(N_{z}+1\right)\times\left( N_z+1\right)}.
\label{jacobiana_henon}
\end{equation}

In addition, the Jacobian matrices for the cases $N_z=1$, $N_z=2$, and $N_z=3$ can not be obtained directly from \eqref{eq:henonfiltrated} and are addressed in the sequel. For $N_z=1$ the system 
\eqref{eq:henonfilt} is rewritten as
\begin{equation}
\left\{\begin{array}{l}
x_1(n+1)=\alpha-\left(c_0 x_1(n)+c_1 x_2(n)\right)^2+\beta x_2(n) \\
x_2(n+1)=x_1(n)
\end{array},\right.
\end{equation}
and its Jacobian matrix is given by
\begin{equation}
\mathbf{J}_{1}\left(\mathbf{x}(n)\right)=\left[\begin{array}{cc}
-2 c_0^2 x_1(n)-2 c_0 c_1 x_2(n) & -2 c_0 c_1 x_1(n)-2 c_1^2 x_2(n)+\beta \\
1 & 0
\end{array}\right]_{2\times2} .
\end{equation}
For $N_{z} = 2$, the system \eqref{eq:henonfiltrated} is rewritten as
\begin{equation}
\left\{\begin{array}{l}
x_1(n+1)=\alpha-(x_3(n))^2 + \beta x_2(n) \\
x_2(n+1)=x_1(n) \\
x_3(n+1)=c_0 \left[\alpha-(x_3(n))^2 + \beta x_2(n)\right] + c_1x_1(n) + c_2x_2(n)
\end{array} \right. ,
\end{equation}
and its Jacobian matrix is given by
\begin{equation}
\mathbf{J}_2\left(\mathbf{x}(n)\right)=\left[\begin{array}{ccc}
0 & \beta & -2x_3(n)\\
1 & 0 & 0 \\
c_1 & c_0 \beta+c_2 & -2c_0 x_3(n)
\end{array}\right]_{3\times3} .
\end{equation}
Finally, for $N_{z} = 3$, the system \eqref{eq:henonfiltrated} is rewritten as
\begin{equation}
\left\{\begin{array}{l}
x_1(n+1)=\alpha-(x_3(n))^2 + \beta x_2(n) \\
x_2(n+1)=x_1(n) \\
x_3(n+1)=c_0 \left[\alpha-(x_3(n))^2 + \beta x_2(n)\right] + c_1x_1(n) + c_2x_2(n) + c_{3}x_{4}(n)\\
x_4(n+1)=x_{2}(n)
\end{array} \right. ,
\end{equation}
and its Jacobian matrix given by

\begin{equation}
\mathbf{J}_3\left(\mathbf{x}(n)\right)=\left[\begin{array}{cccc}
0 & \beta & -2x_3(n) & 0\\
1 & 0 & 0 & 0\\
c_1 & c_0 \beta+c_2 & -2c_0 x_3(n) & c_3\\
0 & 1 & 0 & 0
\end{array}\right]_{4\times4} .
\end{equation}

\section*{Acknowledgments}
This study was financed by CNPq-Brazil (grants 140081/2022-4, 303826/2022-3, 311039/2019-7, and 404081/2023-1), FAPESP (grant 2021/02063-6), and CAPES-Brazil (Finance Code 001).


\end{document}